\documentclass{article}
\usepackage[T1]{fontenc}
\usepackage{wrapfig}
\usepackage[portuges,english]{babel}
\usepackage{amssymb,latexsym,amsmath,color,mathrsfs,ifsym,graphics,stmaryrd} 
\usepackage[colorlinks,linkcolor=blue,urlcolor=blue,citecolor=black,
plainpages=false,pdfpagelabels,breaklinks]{hyperref}
\usepackage[all]{xy}
\usepackage{graphicx}

\title{{\sc The (Quantum) Measurement\\ Problem in Classical Mechanics}}

\author{{\sc Christian de Ronde}}
\date{}

\usepackage[margin=2.1 cm]{geometry}

\begin{document}

\bibliographystyle{plain}
\maketitle

\begin{center}
\begin{small}
Philosophy Institute Dr. A. Korn (UBA-CONICET)\\
Engineering Institute (UNAJ)\\
Center Leo Apostel for Interdisciplinary Studies (Vrije Universiteit Brussel)\\
\end{small}
\end{center}

\bigskip

\begin{abstract}
\noindent In this work we analyze the deep link between the 20th Century positivist re-foundation of physics and the famous measurement problem of quantum mechanics. We attempt to show why this is not an ``obvious'' nor ``self evident'' problem for the theory of quanta, but rather a direct consequence of the empirical-positivist understanding of physical theories when applied to the orthodox quantum formalism. In contraposition, we discuss a representational realist account of both physical `theories' and `measurement' which goes back to the works of Einstein, Heisenberg and Pauli. After presenting a critical analysis of Bohr's definitions of `measurement' we continue to discuss the way in which several contemporary approaches to QM ---such as decoherence, modal interpretations and QBism--- remain committed to Bohr's general methodology. Finally, in order to expose the many inconsistencies present within the (empirical-positivist) presuppositions responsible for creating the quantum measurement problem, we show how through these same set of presuppositions it is easy to derive a completely analogous paradox for the case of classical mechanics. 
\end{abstract}
\begin{small}

{\bf Keywords:} {\em Measurement problem, positivism, Bohr, quantum mechanics, classical mechanics.}
\end{small}

\newtheorem{theo}{Theorem}[section]
\newtheorem{definition}[theo]{Definition}
\newtheorem{lem}[theo]{Lemma}
\newtheorem{met}[theo]{Method}
\newtheorem{prop}[theo]{Proposition}
\newtheorem{coro}[theo]{Corollary}
\newtheorem{exam}[theo]{Example}
\newtheorem{rema}[theo]{Remark}{\hspace*{4mm}}
\newtheorem{example}[theo]{Example}
\newcommand{\proof}{\noindent {\em Proof:\/}{\hspace*{4mm}}}
\newcommand{\qed}{\hfill$\Box$}
\newcommand{\ninv}{\mathord{\sim}} 
\newtheorem{postulate}[theo]{Postulate}

\bigskip

\bigskip

\bigskip

\bigskip

\section{The Measurement Problem in Quantum Mechanics}

Still today, the measurement problem continues to be considered by many as {\it the} most important obstacle for a proper understanding of Quantum Mechanics (QM). The problem is strictly related to the  the existence of superposed states within the quantum formalism and the seemingly contradicting fact that we never actually observe such superpositions in the lab. As argued by orthodoxy, instead of quantum superpositions we only observe single outcomes after a quantum measurement is performed. This apparent contradiction between the mathematical formalism ---which in most cases provides an account of physical situations in terms of quantum superpositions--- and observation gives rise to the famous  measurement problem of QM, which can be stated in the following terms:

\smallskip 
\smallskip 

\noindent {\it {\bf Quantum Measurement Problem (QMP):} Given a specific basis (or context), QM describes mathematically a quantum state in terms of a superposition of, in general, multiple states. Since the evolution described by QM allows us to predict that the quantum system will get entangled with the apparatus and thus its pointer positions will also become a superposition,\footnote{Given a quantum system represented by a superposition of more than one term, $\sum c_i | \alpha_i \rangle$, when in contact with an apparatus ready to measure, $|R_0 \rangle$, QM predicts that system and apparatus will become ``entangled'' in such a way that the final `system + apparatus' will be described by  $\sum c_i | \alpha_i \rangle  |R_i \rangle$. Thus, as a consequence of the quantum evolution, the pointers have also become ---like the original quantum system--- a superposition of pointers $\sum c_i |R_i \rangle$. This is why the measurement problem can be stated as a problem only in the case the original quantum state is described by a superposition of more than one term.} the question is why do we observe a single outcome instead of a superposition of them?}

\smallskip 
\smallskip 

\noindent An obvious way out of the QMP is to take an anti-realist viewpoint and  simply dissolve it. By refusing to attach any representational reference to quantum superpositions anti-realists escape the paradox right from the start. This path is clearly exemplified by Chris Fuchs and Asher Peres \cite[p. 70]{FuchsPeres00} who argue that ``quantum theory does not describe physical reality. What it does is provide an algorithm for computing probabilities for the macroscopic events (`detector clicks') that are the consequences of our experimental interventions.'' On the opposite corner, for the realist, who attempts to understand what is going on beyond measurement, things are quite different. As recently pointed out by Matthias Egg \cite{Egg18}: ``many metaphysicians (and some physicists) consider the abandonment of realism too high a price to pay and therefore insist that the measurement problem calls for a (realistic) solution rather than a dissolution along non-realist lines.'' Accordingly, Egg argues that there exist only two options: ``either [...] to modify the physics (such as theories with additional variables or spontaneous collapses) or drastically inflate the empirically inaccessible content of reality (such as many-worlds interpretations).'' As we shall discuss in detail, these ---supposedly--- realist approaches to QM assume uncritically the naive empiricist standpoint ---which goes back to the positivist understanding of theories--- according to which observations can be considered as {\it givens} of experience, {\it prior} and independent of theoretical representation. Consequently, Egg's map restricts the possibilities of analysis to the acceptance of the just mentioned naive empiricist presupposition. Taking distance from both anti-realism and empiricist versions of realism, we will argue that a truly realist account of QM, necessarily implies a critical reconsideration of the role played by observation within physical theories. Against both naive realism and naive empiricism, observations in physics cannot be regarded as ``common sense'' {\it givens}. On the very contrary, their role is one derived from theories themselves since ---as stressed repeatedly by both Einstein, Heisenberg and Pauli--- it is only the theory which decides what can be observed. And it is only through the creation and development of adequate physical concepts that we will be finally able to understand quantum phenomena. 

The QMP takes for granted a series of empirical-positivist presuppositions which fail to address observations in a critical (realist) manner. In this work we attempt to expose its untenability showing that, taking as a  standpoint the empirical-positivist presuppositions, one can also derive an analogous `measurement problem' for classical mechanics. The article is organized as follows. In section 2 we consider the main scheme and cornerstones of the 20th positivist re-foundation of physics. In section 3 we revisit the deep relation between the QMP and the empirical-positivist understanding of physical theories. Section 4 restates in a contemporary manner the original Greek meaning of realism within physics as strictly founded on the possibility of a theoretical (formal-conceptual) representation of {\it physis}. In section 5, in order to expose the untenability of the QMP for realist perspectives, we show how an analogous problem can be also derived in the case of classical mechanics.

\section{The 20th Century Positivist Re-Foundation of Physics}

In the 17th Century metaphysics had escaped the constraints of experience and advanced wildly as a ``mad dog'' producing all sorts of amazing stories about existence and reality. There was no limit nor constrain for dogmatic metaphysics which debated without any doubt, not only about about the existence of God, but also about the properties of angels. In his fight against dogmatic metaphysics, the physicist and philosopher Immanuel Kant went back to the ---apparently--- more humble Sophistic viewpoint. On the one hand, he recognized that the standpoint of science needed to be human experience; on the other, he accepted the finitude of man, and consequently, the impossibility to access the infinite things-in-themselves ---i.e., objects as they are, independent of human perception. Kant reconfigured the foundation of knowledge itself grounding his new philosophy in a {\it transcendental subject} that captured through the table of (Aristotelian) {\it categories} and the {\it forms of intuition} (i.e., Newtonian space and time), the way in which we, humans, experience phenomena. From this standpoint, he was able to develop a new metaphysical architectonic in which the (transcendental) subject played the most fundamental role, providing the conditions of possibility for objective experience itself; i.e., the experience about objects. In his {\it Critique of Pure Reason} he was then able to explain how Newtonian physics could be understood as providing {\it objective knowledge} about the phenomena observed by (empirical) subjects. By understanding the limits of human experience metaphysics would finally follow the secure path of science showing, at the same time, how (scientific) knowledge is possible. Unfortunately, very soon, the {\it a priori} forms of intuition (i.e., Newtonian absolute space and time) were taken as the fundamental unmovable cornerstones of all possible experience, restricting in this way all access to new phenomena and turning Kantian metaphysics itself into a new dogma. 

Already in the mid 19th Century, the cracks in the structure of the Kantian building had begun to become visible. The categories and forms of intuition put forward by Kant had turned into the same he had striven to attack in the metaphysics of his time, dogmatic and unquestionable elements of thought. As noted by van Fraassen \cite[p. 2]{VF02}: ``Kant exposed the illusions of Reason, the way in which reason overreaches itself in traditional metaphysics, and the limits of what can be achieved within the limits of reason alone. But on one hand Kant's arguments were not faultless, and on the other there was a positive part to Kant's project that, in his successors, engaged a new metaphysics. About a century later the widespread rebellions against the Idealist tradition expressed the complaint that Reason had returned to its cherished Illusions, if perhaps in different ways.'' In the mid-19th Century, Ernst Mach ---another physicist and philosopher--- produced then the most radical deconstruction of the main concepts of classical mechanics ---which were also founding the Kantian scheme. Mach was able to derive a new positivist understanding of physics in which {\it sensations} would play the most fundamental role. His investigations led him to the conclusion that science is nothing but the systematic and synoptical recording of data of experience. In his famous book, {\it Analysis of Sensations}, he wrote:
\begin{quotation}
\noindent {\small``Nature consists of the elements given by the senses.
Primitive man first takes out of them certain complexes of these
elements that present themselves with a certain stability and are
most important to him. The first and oldest words are names for
`things'. [...] The sensations are no `symbols of things'. On the
contrary the `thing' is a mental symbol for a sensation-complex of
relative stability. Not the things, the bodies, but colors, sounds,
pressures, times (what we usually call sensations) are the true
elements of the world.'' \cite{Mach}}
\end{quotation}
\noindent  Mach concluded that primary sensations constitute the ultimate building blocks of science, inferring at the same time that scientific concepts are only admissible if they can be defined in terms of sensations. In close analogy to Darwinistic ideas, Mach conceived the evolution of knowledge in physical theories as a process of ``struggle for life'' and ``survival of the fittest''. Although Mach had been himself a neo-Kantian, within his new positivist conception of science, he stated that we should reject every {\it a priori} element in the constitution of our knowledge about things. This move took experience and observation back to its naive pre-Kantian form. Scientific propositions should be empirically verifiable and science would be then nothing but a conceptual reflection of facts ---as provided by sensations.

Mach's pre-Kantian empiricist standpoint allowed him not only to criticize the physical concepts of classical mechanics, but also to produce a complete reformulation of the meaning and applicability of physical theories. Furthermore, Mach deconstruction of the {\it a priori} (classical) concepts of `space', `time', `substance', `causality', etc., allowed the next generation of physicists ---Einstein, Bohr, Heisenberg, Pauli and many others---, to use this new freedom in order to begin to conceive not only a new experience ---beyond classical notions--- but also new physical theories. It is well known that the philosophical ideas of Mach had a great influence in the development of special and general relativity. Albert Einstein indicated on several occasions the relations of his own ideas to those of Mach, in whom he recognized a guide. But the importance of Mach's thought should not be underestimated in relation to the development of quantum theory. Heisenberg, used the Machian principle according to which only observables should be considered within a theory in order to develop matrix mechanics ---escaping right from the start the (classical) question regarding `the trajectories of particles'.\footnote{As noticed by Arthur Fine \cite[p. 1195]{PS}: ``Heisenberg's seminal paper of 1925 is prefaced by the following abstract, announcing, in effect, his philosophical stance: `In this paper an attempt will be made to obtain bases for a quantum-theoretical mechanics based exclusively on relations between quantities observable in principle'.''} At the beginning of the 20th Century, the Machian epistemological principle had finally broken the chains of classical Newtonian physics and Kant's metaphysics. A new quantum experience was disclosed, a new region of thought had been created.

Concomitant with the creation of relativity and QM, a new generation of positivists like Rudolph Carnap, Otto Neurath and Hans Hahn congregated in what they called `the Vienna Circle'. In their famous and influential manifesto \cite{VC} they wrote the following: ``In science there are no `depths'; there is surface everywhere: all experience forms a complex network, which cannot always be surveyed and, can often be grasped only in parts. Everything is accessible to man; and man is the measure of all things. Here is an affinity with the Sophists, not with the Platonists; with the Epicureans, not with the Pythagoreans; with all those who stand for earthly being and the here and now.''  Metaphysics was now understood as a dogmatic discourse about unobservable entities. According to them, science should focus in ``statements as they are made by empirical science; their meaning can be determined by logical analysis or, more precisely, through reduction to the simplest statements about the empirically given.'' This new positivism sedimented Mach's pre-Kantian understanding of observation. In turn, this naive standpoint allowed them to draw a main distinction ---that would guide the science of the new century--- between ``empirical terms'', the empirically ``given'' in physical theories and ``theoretical terms'', their translation into simple statements. An important consequence of this naive empiricist standpoint is that physical concepts become only supplementary elements, parts of a narrative that could be added, or not, to an already empirically adequate theory. When a physical phenomenon is understood as a self-evident {\it given} ---independent of physical concepts and metaphysical presuppositions---, empirical terms configure an ``objective'' set of data which can be directly related to a formal scheme. Objective here does not relate to a conceptual {\it moment of unity} ---like in the Kantian scheme--- but instead, to something quite undefined like a ``good'' or ``honest'' observation made by subject. 
\begin{center} 
{\it Empirical Observable Data --------------- Theoretical Terms

\smallskip 
\smallskip 

(Supplementary `Interpretation' or `Fictional Narrative')}
\end{center}



\noindent Since an empirically adequate theory is already able to predict the observations it talks about, there seems to be no essential need to add a ``story'' or ``interpretation'' about the external reality outside observation. The role of concepts becomes then completely accessory to those needing to believe in narratives about how the world could be according to the theory. But as remarked by van Fraassen  \cite[p. 242]{VF91}: ``However we may answer these questions, believing in the theory being true or false is something of a different level.'' According to the empiricist viewpoint, the world is unproblematically described in terms of our ``common sense'' access to experience and an adequate empirical theory can perfectly account for experiments without the need of any interpretation. 

The project of articulating the empirical-formal relation through the distinction between {\it theoretical terms} and {\it observational terms} never accomplished the promise of justifying their independence ---specially with respect to the categorical or metaphysical definition of concepts. The fundamental reason had already been discussed in Kant's {\it Critique of Pure Reason}: the observation of a phenomenon cannot be considered without previously taking into account a categorical-metaphysical scheme. The description of phenomena always necessarily presupposes  ---implicitly or explicitly--- certain metaphysical principles. `Identity' or `non-contradiction' are not `things' that we see walking in the street, but rather the very {\it conditions of possibility} of classical experience itself. We never observe these principles (e.g. causality); rather, we presuppose them in order to make sense of phenomena. This categorical systematization, allowing for a theoretical-conceptual representation, is in itself metaphysical. As the philosopher from K\"onigsberg would have said, it is the representational framework of the transcendental subject, articulating categories and forms of intuition, that which allows for an objective experience of the empirical subject.\footnote{In the context of analytic philosophy, this metaphysical aspect of observation was re-discovered by Hanson in the mid 20th Century \cite{Hanson}. However, regardless of the deep conclusion, the main positivist scheme of thought remained as a standpoint of analysis, taking also for granted a naive empiricist standpoint.} In fact, this was according to Einstein the really significant  philosophical achievement of Kant:  
\begin{quotation}
\noindent {\small``From Hume Kant had learned that there are concepts (as, for example, that of causal connection), which play a dominating role in our thinking, and which, nevertheless, can not be deduced by means of a logical process from the empirically given (a fact which several empiricists recognize, it is true, but seem always again to forget). What justifies the use of such concepts? Suppose he had replied in this sense: Thinking is necessary in order to understand the empirically given, {\it and concepts and `categories' are necessary as indispensable elements of thinking.}'' \cite[p. 678]{Einstein65} (emphasis in the original)}
\end{quotation} 

\noindent Maybe due to the general failure of the program \cite{Hempel}, it is difficult to find today anyone willing to call herself a positivist. However, regardless of its failure, the basic ideas present in early positivism have reached our days and ---more importantly--- continue to configure the problems we discuss in the present. The limits of positivism itself are of course difficult to encapsulate, but there are four main cornerstones which might be regarded to conform the main positivist program. These pillars have played an essential role in the deep 20th Century postmodern re-foundation of physics. 
\begin{enumerate}
{\bf  \item[I.] Naive Empiricism:} observation is a self evident {\it given} of ``common sense'' experience. 
{\bf \item[II.] Physics as an Economy of Experience:} physical theories are mathematical machineries (algorithmic formalisms) which produce predictions about observable measurement outcomes. 
{\bf \item[III.] Anti-Metaphysics:} metaphysics, understood as an interpretation or narrative about {\it unobservable entities}, is not essentially required within empirically adequate theories. 
{\bf  \item[IV.] Intersubjective Justification:} the {\it intersubjective} communication of theoretically predicted observable data between the members of a scientific community allows to regard a theory as ``objective''. 
\end{enumerate}

All `isms' in science imply a specific {\it praxis} defined by the field of its problems. Scientific problems are not independent of the understanding of what the scientific enterprise is really about. In this respect, the influence of the just mentioned pillars ---regardless of the fact positivism has become a quite unpopular term--- continue to play still today a kernel role within the discussions and debates that conform the specialized physical and philosophical literature about QM. It is important to realize that in order to be part of a program, it doesn't really matter if you truly believe in its cornerstones; what matters is what you do. If you follow a program, then ---quite regardless of your personal beliefs--- you must engage with its problems and discuss what is important for the program. In this respect, one of the main problems introduced by positivism within the theory of quanta has been the famous measurement problem.

\section{The Positivist Origin of the Quantum Measurement Problem}
 
As we discussed above, positivist ideas were deeply influential for the creation and development of QM. In 1925, Werner Heisenberg was finally able to find the key to develop the closed mathematical formalism of matrix mechanics through the Machian positivist standpoint according to which only observable quantities are to be considered within a theory. Positivism allowed him to abandon the focus on the classical trajectories of particles and restate the problem from a completely different angle. One year later, following the work by Louis de Broglie on matter waves, Erwin Schr\"odinger proposed a wave equation which introduced the famous quantum wave function, $\Psi$. Regardless of the initial expectations to restore the classical space-time representation, it was soon realized that the equation was written in {\it configuration space}, precluding its classical understanding in terms of a space-time 3-dimensional wave. That same year Max Born presented his probabilistic interpretation of $\Psi$, which only made reference to the prediction of `clicks' in detectors consequence of elementary particles. This interpretation ---which Einstein strongly rejected---, also made implicit use of the positivist {\it Zeitgeist} according to which theories, instead of describing a real states of affairs, had to be understood as ``tools'' for predicting observations. As remarked by Osnaghi et al. in \cite{OsnaghiFreitasFreire09}: ``During the 1920s and 1930s, the ideas which were to be identified with the `orthodox view' of quantum mechanics became quite popular. The positivist flavor of the approach developed by Heisenberg, Jordan, Born and Pauli was not only in tune with the cultural climate of continental Europe between the two wars, but was also well suited to cope with the change of paradigm that atomic phenomena seemed to demand.'' Regardless of the fact that QM had no clear representational reference, in this context ---following the positivist understanding of physical theories--- the focus was pointed to the need of securing observability. 

After the theoretical developments of Heisenberg and Schr\"odinger, Paul Dirac, a young English mathematician, attempted to provide a sound axiomatic mathematical presentation of the new theory. In 1930 he presented his book, {\it The Principles of Quantum Mechanics}, where from an explicit positivist perspective he stressed that it is ``important to remember that science is concerned only with observable things and that we can observe an object only by letting it interact with some outside influence. An act of observation is thus necessarily accompanied by some disturbance of the object observed.'' Following Bohr's teachings, he also remarked that \cite[pp. 3-4]{Dirac74}: ``we have to assume that {\it there is a limit to the finiteness of our powers of observation and the smallness of the accompanying disturbance ---a limit which is inherent in the nature of things and can never be surpassed by improved technique or increased skill on the part of the observer.}'' Dirac was maybe the first to realize the importance of the superposition principle which implied the existence of strange quantum superpositions. Dirac realized that this introduced a serious obstacle for his positivist reading. Experiment simply did not provide access to the many superposed states, but only to a single outcome. In order to bridge this gap, making use of an example of polarized photons, Dirac introduced for the first time the now famous ``collapse'' of the quantum wave function: 
\begin{quotation}
\noindent {\small ``When we make the photon meet a tourmaline crystal, we are subjecting it to an observation. We are observing wither it is polarized parallel or perpendicular to the optic axis. The effect of making this observation is to force the photon entirely into the state of parallel or entirely into the state of perpendicular polarization. It has to make a sudden jump from being partly in each of these two states to being entirely in one or the other of them. Which of the two states it will jump cannot be predicted, but is governed only by probability laws.'' \cite[p. 7]{Dirac74}} 
\end{quotation}
Even though the explanation did not succeed in providing a consistent `picture' of what was going on, Dirac recalled his readers that ``the main object of physical science is not the provision of pictures, but the formulation of laws governing phenomena and the application of these laws to the discovery of phenomena. If a picture exists, so much the better; but whether a picture exists of not is a matter of only secondary importance.'' An empirically adequate theory might possess an `interpretation', but this is not essential, the scheme only  ``becomes a precise physical theory when all the axioms and rules of manipulation governing the mathematical quantities are specified and when in addition certain laws are laid down connecting physical facts with the mathematical formalism.''

Two years later, in 1932, an Hungarian mathematician called John von Neumann published his famous {\it Mathematische Grundlagen der Quantenmechanik} where he attempted to provide an even more rigorous mathematical formulation of QM.\footnote{Von Neumann was strongly influenced by Hilbert regarding the axiomatization program. In a joint paper by Hilbert, Nordheim and von Neumann from 1926 (translated by Redei and St\"oltzner) they claim: ``In physics the axiomatic procedure alluded to above is not followed closely, however; here and as a rule the way to set up a new theory is the following. One typically conjectures the analytic machinery before one has set up a complete system of axioms, and then one gets to setting up the basic physical relations only through the interpretation of the formalism. It is difficult to understand such a theory if these two things, the formalism and its physical interpretation, are not kept sharply apart. This separation should be performed here as clearly as possible although, corresponding to the current status of the theory, we do not want yet to establish a complete axiomatics. What however is uniquely determined, is the analytic machinery which ---as a purely mathematical entity--- cannot be altered. What can be modified ---and is likely to be modified in the future--- is the physical interpretation, which contains a certain freedom and arbitrariness.''} In this case, the need of an axiomatization of the theory might be easily linked to von Neumann's empiricist understanding of the method of the physical sciences (see also \cite{Bueno16}).
\begin{quotation}
\noindent {\small ``To begin, we must emphasize a statement which I am sure you have heard before, but which must be repeated again and again. It is that the sciences do not try to explain, they hardly ever try to interpret, they mainly make models. By a model is meant a mathematical construct which, with the addition of some verbal interpretations describes observed phenomena. The justification of such a mathematical construct is solely and precisely that it is expected to work ---that is correctly to describe phenomena from a reasonably wide area.''  \cite[p. 9]{RedeiStoltzner}}\end{quotation} 
This empiricist standpoint assumed by both Dirac and von Neumann seemed to confront the existence of superpositions in QM which did not seem to describe what was being actually observed. But after Dirac had introduced the ``collapse'', von Neumann \cite[p. 214]{VN} was ready to transform it into a {\it postulate} of the theory itself: ``Therefore, if the system is initially found in a state in which the values of $\mathcal{R}$ cannot be predicted with certainty, then this state is transformed by a measurement $M$ of $\mathcal{R}$ into another state: namely, into one in which the value of $\mathcal{R}$ is uniquely determined. Moreover, the new state, in which $M$ places the system, depends not only on the arrangement of $M$, but also on the result of $M$ (which could not be predicted causally in the original state) ---because the value of $\mathcal{R}$ in the new state must actually be equal to this $M$-result.''  The {\it projection postulate} secured the observation of `clicks' in detectors and `spots' in photographic plates, but this addition came with a great cost. Like a Troyan horse, the postulate hosted an unwanted guest. Apart from the unitary deterministic evolution guided by Schr\"odinger's equation of motion, applicable when we do not observe what is going on, there was now a new non-unitary indeterministic evolution that took place every time we measured (or observed) a quantum state. Measurement in QM seemed to be responsible in producing a real physical process which was not represented by the theory. However, from a positivist viewpoint which embraced the sophistic idea that `man is the measure of all things' this ideas did not seem so strange after all. 

This new ``quantum jump'', different to those made popular by Bohr,\footnote{It is important to remark however that the jumps discussed by Bohr took place within the atomic representation. Electrons jumped from one orbit to the next without any clear explanation. So while Bohr's jumps made reference to a limit of the representation within the theory, Dirac's jumps made reference to the shift from the mathematical representation of superpositions to the empirical realm of observation.} was between the mathematical formulation and {\it hic et nunc} observations made by subjects. The need of introducing an agent capable of observing, implied the breakdown of the possibility of an objective (subject-independent) theoretical representation. Without any success in the physics community, Einstein and Schr\"odinger would raise their voice against this unsatisfactory evolution of the theory commanded by what could be seen as a silent alliance between Bohr and positivists. As Einstein \cite[p. 7]{OsnaghiFreitasFreire09} would mention to Everett some years later, he ``could not believe that  a mouse could bring about drastic changes in the universe simply by looking at it.'' Also Schr\"odinger criticized explicitly, not only the collapse, but also the explicit subjectivity involved within the process: 
\begin{quotation}
\noindent {\small``But jokes apart, I shall not waste the time by tritely ridiculing the attitude that the state-vector (or wave function) undergoes an abrupt change, when `I' choose to inspect a registering tape. (Another person does not inspect it, hence for him no change occurs.) The orthodox school wards off such insulting smiles by calling us to order: would we at last take notice of the fact that according to them the wave function does not indicate the state of the physical object but its relation to the subject; this relation depends on the knowledge the subject has acquired, which may differ for different subjects, and so must the wave function.'' \cite[p. 9]{OsnaghiFreitasFreire09}} 
\end{quotation}

Unfortunately, since the positivist {\it Zeitgeist} had already captured the understanding of physical theories, and the critical remarks by Einstein and Schr\"odinger were not seriously considered. The orthodox textbook reading became the following: quantum states behave in a {\it deterministic} manner when there is no measurement, but there is also an {\it indeterministic} evolution which takes place each time we attempt to observe what is really going on. As noticed by Dennis Dieks \cite[p. 120]{Dieks10}: ``Collapses constitute a process of evolution that conflicts with the evolution governed by the Schr\"{o}dinger equation. And this raises the question of exactly when during the measurement process such a collapse could take place or, in other words, of when the Schr\"{o}dinger equation is suspended. This question has become very urgent in the last couple of decades, during which sophisticated experiments have clearly demonstrated that in interaction processes on the sub-microscopic, microscopic and mesoscopic scales collapses are never encountered.'' In the last decades, the experimental research seems to confirm there is nothing like a ``real collapse'' taking place when measurements are performed. In a more recent paper, Dieks \cite{Dieks18} acknowledges the fact that: ``The evidence against collapses has not yet affected the textbook tradition, which has not questioned the status of collapses as a mechanism of evolution alongside unitary Schr\"odinger dynamics.''   

\smallskip 

In order to disentangle the mess created by the application of the positivist pillars to QM, it might be wise to go back to the representational realist understanding of physics, and in particular, to the writings on these matters of Einstein, Heisenberg and Pauli. 

\section{Realism in Physics: Theoretical Representation and Experience}

The origin of physics is intrinsically related to the ancient Greek attempt to study and understand, in theoretical terms, what physicists and philosophers called {\it physis} ---later on translated as reality or nature. Physics begins by presupposing the existence of {\it physis}, not only as the fundament  of existence itself but also as a {\it moment of unity} within thought. This is the fundamental pillar of physical (or realistic) thought. Physical theories have always attempted to provide a {\it representation} of reality (or {\it physis}). 
\begin{center}
{\it Physical Theory / Formal-Conceptual Representation} \ \  -------- \ \ {\it Physis / Nature / Reality} 
\end{center}
\noindent Physics and realism go hand in hand, simply because physicists are mainly interested in reality, which is just a translation of {\it physis} \cite{Cordero14}. One of the main presuppositions of physical realism is that {\it physis} is not chaotic, it has an internal order, a {\it logos}. And theories are able ---in some way--- to express the {\it logos} of {\it physis}, making true scientific knowledge is possible. Physics and realism take as a basic standpoint that there exists a {\it relation} between theory and reality. But this does not imply the general widespread the naive (pre-Kantian) claim according to which the relation between theory and reality should be regarded as a one-to-one {\it correspondence relation}..\footnote{This idea, according to which the theory describes {\it reality as it is} can be only sustained by extremely naive forms of realism such as scientific realism; a formulation created by empiricists much closer to the positivists pillars than to the {\it praxis} of either science or realism.} Of course, one can think of many different more interesting {\it relations} between theory and {\it physis} \cite{deRonde14, deRonde16}. 

But the characterization of realism does not end here, it continues with a particular account of the relation between theoretical representation and experience. For realists ---contrary to anti-realists--- theoretical representation comes always first, experience is necessarily second. Experience is derived from theoretical representation itself which is conformed by an interrelated formal-conceptual framework. As Einstein would tell to a very young Heisenberg: ``It is only the theory which decides what can be observed.'' Theoretical representation is first, experience and perception are necessarily second. This marks the kernel point of disagreement between empiricism ---from which positivism, instrumentalism and even scientific realism were developed--- and realism. According to the latter, willingly or not, we physicists, are always producing our praxis {\it within} a specific representation. And in this respect, there is no such thing as `internal' and `external' realities. Physical analysis takes always as a standpoint the theoretical representation of a state of affairs. There is nothing ``outside'' the gates of representation. 

\smallskip 
\smallskip 

\noindent {\bf Realism:} The working presupposition that there exists a {\it relation} between {\it theory} and {\it physis} and that it is possible to develop theoretical representations which are independent of subjects or any preferred perspective.

\smallskip
\smallskip

\noindent Representation is not only mathematical or formal, it is also conceptual or metaphysical. And this marks the second point of distinction between realism and anti-realism, namely, the meaning and role played by metaphysics ---which is not that presented by positivists.\footnote{For a detailed discussion of the orthodox role played by metaphysics in philosophy of physics we refer to \cite{Arenhart19}.} While for anti-realists metaphysics is a mere discourse or story about the un-observable realm, a narrative that might be added to an empirically adequate theory; for realists, metaphysics is the systematic creation of a relational net of concepts which constitute the condition of possibility for thought and experience themselves. As remarked by Einstein:
\begin{quotation}
\noindent {\small ``I dislike the basic positivistic attitude, which from my point of view is untenable, and which seems to me to come to the same thing as Berkeley's principle, {\it esse est percipi.} `Being' is always something which is mentally constructed by us, that is, something which we freely posit (in the logical sense). The justification of such constructs does not lie in their derivation from what is given by the senses. Such a type of derivation (in the sense of logical deducibility) is nowhere to be had, not even in the domain of pre-scientific thinking. The justification of the constructs, which represent `reality' for us, lies alone in their quality of making intelligible what is sensorily given.'' \cite[p. 669]{Einstein65}}
\end{quotation} 
It is only through such generalizations that thought becomes possible, that we can even think and imagine an experience which was never actually observed. As Heisenberg \cite[p. 264]{Heis73} made the point: ``The history of physics is not only a sequence of experimental discoveries and observations, followed by their mathematical description; it is also a history of concepts. For an understanding of the phenomena the first condition is the introduction of adequate concepts. Only with the help of correct concepts can we really know what has been observed.'' But this distinction regarding the meaning of metaphysics is also a distinction regarding the meaning and role played by concepts. While for anti-realists concepts {\it refer} to `things' in the world, in the case of realism concepts are always related to other concepts. Concepts are intrinsically relational. Concepts cannot exist alone. A concept is always part of a conceptual net, a whole in which each element is supported by its neighbor ---and viceversa. As remarked by Heisenberg  \cite[p. 94]{Heis58}: ``the connection between the different concepts in the system is so close that one could generally not change any one of the concepts without destroying the whole system.'' For example, the notion of `object' (e.g., a `table') implies a systematic architectonic supported by specific logical and metaphysical principles, namely, the principles of existence, non-contradiction and identity. If you leave out any of these concepts, the notion immediately looses its whole meaning. But the notion of object has meaning only when also related to other notions ---such as, e.g., the notion of space, the notion of time, etc. They all connect in such a way that we can think ---in a specific manner--- about physical reality. This is the reason why the discovery a new field of experience implies always the creation of new conceptual frameworks and mathematical formalism. As Heisenberg \cite{Boku04} explains: ``the transition in science from previously investigated fields of experience to new ones will never consist simply of the application of already known laws to these new fields. On the contrary, a really new field of experience will always lead to the crystallization of a new system of scientific concepts and laws''. In physics, theoretical representations are not only metaphysical or conceptual, they are also formal or mathematical. Only together, a theory is capable of producing a qualitative and quantitative representation and understanding of experience. While the conceptual part provides a {\it qualitative} type of understanding, the formal part provides a {\it quantitative} understanding. In order to conform a whole, both parts must be linked in a structural manner. Metaphysical concepts must be consistently and systematically related to mathematical notions ---and viceversa. It is this relation which provides the missing link between mathematics and metaphysics. Once again, the notion of `object' provides a very good example of such intrinsic relation since the principles of existence, non-contradiction and identity play not only a metaphysical role, but also a logical one. In fact, they determine classical (Aristotelian) logic itself, which in turn, is also related to a specific mathematical formalism which is consistent with these principles (for a detailed discussion see \cite{deRondeBontems11, deRondeFreytesDomenech}). 
 
Another essential point of physical theoretical representations is the construction of a scheme in which their reference becomes completely independent of the particular viewpoints taken by (empirical) subjects. Agents and their observations can play no essential role within the physical representation of a state of affairs. As Einstein \cite[p. 175]{Dieks88a} made the point: ``[...] it is the purpose of theoretical physics to achieve understanding of physical reality which exists independently of the observer, and for which the distinction between `direct observable' and `not directly observable' has no ontological significance.'' While the formal aspect allowing to detach theoretical representation from (empirical) subjects is provided via the {\it invariance} of the mathematical formalism, the conceptual component is articulated by the Kantian notion of {\it objectivity}. The notions of invariance and objectivity are intrinsically related; one being the counterpart of the other (see for a detailed analysis \cite{deRondeMassri17}). These two definitions, combined, allow physics to represent things as a {\it moment of unity} independently of subjects and their particular observations. Physics does not deny the existence of (empirical) subjects, it simply considers them {\it within} Nature. The scientific enterprise considers subjects, not more nor less but {\it as} important as any other existent. This marks a third point of disagreement between realists and anti-realists. While for realists scientific analysis begins always from a theoretical standpoint, from the {\it physis}-perspective; for anti-realists, assuming the {\it relative} perspective of an individual subject, the analysis begins from ``common sense'' observation and perception. While the latter argues that the observation of subjects (or agents) provide the basic elements from which theories are built, realists claim that the systematic understanding of experience can be only produced through the creation of a general theoretical (mathematical and metaphysical) framework. Once again, as Heisenberg makes clear: 
\begin{quotation}
\noindent {\small  ```Understanding' probably means nothing more than having whatever ideas and concepts are needed to recognize that a great many different phenomena are part of coherent whole. Our mind becomes less puzzled once we have recognized that a special, apparently confused situation is merely a special case of something wider, that as a result it can be formulated much more simply. The reduction of a colorful variety of phenomena to a general and simple principle, or, as the Greeks would have put it, the reduction of the many to the one, is precisely what we mean by `understanding'. The ability to predict is often the consequence of understanding, of having the right concepts, but is not identical with `understanding'.'' \cite[p. 63]{Heis71}}
\end{quotation}

However, the most important point of disagreement between realists and anti-realists is marked by the different type of problems ---grounded on essentially different pillars--- they both address in each case. What is important for the realist is not important for the anti-realist, and viceversa. In the context of QM, as we have discussed in \cite{deRonde16, deRonde18a} these problems are simply orthogonal. On the one hand, the anti-realist, assuming that observations in the lab are ``unproblematic'' still has to confront the fact that representation and interpretation seem to be playing an essential role in order to connect empirical observations with the mathematical formalism. On the other hand, the realist, having a mathematical formalism that is capable of operational predictions must still produce a conceptual (subject-independent) representation which is structurally connected to the formalism and is also able to provide an {\it anschaulich} (intuitive) content to the theory. This realist problem is extremely difficult to solve in QM due to the extreme departure which the theory of quanta seems to imply with respect to our (classical) metaphysical representation of (classical) reality in terms of objects moving in space-time. As Heisenberg \cite[p. 3]{Heis58}, made the point: ``the change in the concept of reality manifesting itself in quantum theory is not simply a continuation of the past; it seems to be a real break in the structure of modern science''. Thus, while the anti-realist attempts to build a bridge between the quantum formalism and our ``common sense'' classical or manifest image of the world, the realist is confronted with the need to create a new non-classical conceptual scheme which can be structurally linked to the mathematical formalism of the theory. These different problems are intrinsically related to two very different understandings of `physics', either as providing theoretical (formal-conceptual) representations of reality, or, as a mathematical algorithm capable of predicting the observations of agents. 

Tim Maudlin \cite[p. xii]{Maudlin19} has recently argued that ``physical theories are neither realist or anti-realist [...] It is a person's attitude toward a physical theory that is either realist or antirealist.'' We certainly disagree with this statement. The elements, the structure and the general scheme of realist and anti-realist theories are essentially different. As we have argued above, while an anti-realist theory is a mathematical formalism grounded on observations, a realist theory is constituted by the entrenched relation between a conceptual framework a mathematical formalism and a field of thought-experience. Unlike for the anti-realist, an operational algorithm capable of predictions ---such as QM--- with no consistent conceptual framework  cannot be yet regarded as a closed theory. It still requires a consistent conceptual representation which provides an intuitive ({\it anschaulicht}) access to the experience it talks about.  According to Maudlin: ``The scientific realist maintains that in at least some cases, we have good evidential reasons to accept theories or theoretical claims as true, or approximately true, or on-the-road-to-truth. The scientific antirealist denies this. These attitudes come in degrees: You can be a mild, medium, or strong scientific realist and similarly a mild, medium, or strong scientific antirealist.'' Like many contemporary postmodern characterizations of realism, Maudlin ends up equating realism with the naive belief of an agent. Against this definition, realism must be understood as a working presupposition in which the production of objective (subject-independent) representations is always a main goal to achieve. Being a realist is not defined by a personal belief, it is rather a question which regards the choice of the problems someone chooses to confront. In this respect, being a realist is, so to say, a matter of {\it praxis}.  

\smallskip 
\smallskip 
 
\noindent  {\bf Realist:} Those who assume a {\it praxis} according to which there exists a {\it relation} between {\it theory} and {\it physis}. Those who engage and work on realist problems such as the production of (subject-independent) theoretical representations.

\smallskip 
\smallskip 

\noindent To sum up, while realism implies a perspective from {\it physis} itself, the anti-realist always assumes a perspective relative to a {\it subject}. Realism and anti-realism also imply different pillars from which two completely different fields of problems arise in each case. This distinction implies a different {\it praxis}. To know if you are a realist, you simply have to find out ---regardless of your personal beliefs you might profess in public---- which are the pillars that support the problems you are working on. Wolfgang Pauli, argued in this respect that the main problem that realists confront might imply a major revolution in our way of thinking: 
\begin{quotation}
\noindent {\small ``When the layman says `reality' he usually thinks that he
is speaking about something which is self-evidently known; while to
me it appears to be specifically the most important and extremely
difficult task of our time to work on the elaboration of a new idea
of reality.'' \cite[p. 193]{Laurikainen98}}
\end{quotation}

\section{Measurement: From Theoretical Experience to Actual Observation}

Exposing the fact that theories do not make exclusive reference to observations, theoretical analysis escapes the empiricist standpoint right from the start. A very good example of this is provided by the Kochen-Specker theorem which states ---taking as a standpoint the orthodox mathematical formalism of QM--- that projection operators cannot be related to a {\it Global Binary Valuation} \cite{deRondeMassri18}. This formal result has a deep consequence at the conceptual level, namely, it presents a limit to the possibility of interpreting projection operators in terms of definite valued properties. The theorem goes explicitly beyond the very possibilities of measurement and observation, since it discusses a situation which ---if one accepts the orthodox formalism--- cannot ---by definition--- be subject of measurement (see for a detailed analysis \cite{deRonde19}). This of course does not mean that for physicists (or realists) measurements are unimportant. On the contrary, measurements are an essential way to check out if a theory is capable of {\it expressing} an aspect of reality (or not). At some point, the thought-experience created by a theory in terms of a net of concepts and a mathematical formalism needs to be tested through measurements in the lab. As stressed by Einstein \cite[p. 175]{Dieks88a}, even though observability does not play a role within the theory itself, ``the only decisive factor for the question whether or not to accept a particular physical theory is its empirical success.'' 

Measurement is a way to connect theoretical experience with {\it hic et nunc} observation, it provides the link between an objective theoretical representation and subjective empirical observability. Given a theoretical representation of a state of affairs we can imagine the possible experiences contained within the theory. And this thinking is not restricted by our technical or instrumental capabilities nor by our previous observations, it is only restricted by our theoretical (mathematical and conceptual) schemes of thought. This is the true power of physics. Not only the possibility to escape the here and now creating a thought-experience without the need to actually produce it in the lab, but also, to escape the contemporary technical limitations of our time and advance through thought and representation in our understanding and possibilities of future developments. It is {\it Gedankenexperiments} which show explicitly that, in physics, theoretical experience comes always before actual measurement and observability. In fact, this is exactly what was done in 1935 by both Einstein and Sch\"odinger in their famous EPR and `cat' {\it Gedankenexperiments} where, going beyond the technical restrictions of their epoch, they discussed the experimental and representational consequences of the theory of quanta. It took half a century ---partly due to the unwillingness of positivists, Bohrian and later on instrumentalists--- to actually investigate the existence of quantum superpositions and entanglement; two of the main notions that made possible the ongoing technological revolution of quantum information processing.\footnote{As remarked by Jeffrey Bub \cite{Bub17}, ``[...] it was not until the 1980s that physicists, computer scientists, and cryptographers began to regard the non-local correlations of entangled quantum states as a new kind of non-classical resource that could be exploited, rather than an embarrassment to be explained away.''  The reason behind this shift in attitude towards {\it entanglement} is an interesting one. As Bub continues to explain: ``Most physicists attributed the puzzling features of entangled quantum states to Einstein's inappropriate `detached observer' view of physical theory, and regarded Bohr's reply to the EPR argument (Bohr, 1935) as vindicating the Copenhagen interpretation. This was unfortunate, because the study of entanglement was ignored for thirty years until John Bell's reconsideration of the EPR argument (Bell, 1964).''.} 

Theoretical experience needs to make contact with subjective observation. While the field of thought-experience is strictly limited by the theory itself, observation is a purely subjective conscious action which cannot be theoretically represented. It is `measurement' which stands just in the middle between the theoretical representation of physical experience and the subjective {\it hic et nunc} empirical observation. Measurements, at least for the realist, are conscious actions performed by human subjects which are able to select, reproduce  and understand a specific type of phenomenon. This is of course a very complicated process created by humans which interrelates practical, technical and theoretical knowledge. Anyone attempting to perform a measurement must be able to think about a specific problem, she must be also able to construct a measurement arrangement, she must be capable to analyze what might be going on within the process, and finally, she must be qualified to observe, interpret and understand the phenomenon that actually takes place, {\it hic et nunc}, when the measurement is actually performed. All these requirements imply human capacities and, in particular, consciousness. The table supporting the measurement set up does not understand what complicated process is taking place above itself. Tables and chairs cannot construct a measuring set-up. The chair that stands just beside the table cannot observe a measurement result, and the light entering the lab through the window cannot interpret what is going on. It is only a conscious (empirical) subject (or agent) who is capable of performing a measurement. And these actions have nothing to do with ``common sense''. Measurement is a controlled technical-theoretical activity.

At this point it becomes of outmost importance to clearly distinguish between the theoretical representation of the interactions of `systems' which are already part of a conceptual picture and the empirical observation which requires a conscious agent capable of understanding the theory and interpreting her observation as a measurement.\footnote{It is important not to confuse a `conscious subject' with a `system' described by a theory.} 

\smallskip
\smallskip

\noindent {\it {\sc Theoretical Experience:} The process of interaction and evolution of a state of affairs represented in theoretical terms via the consistent conjunction of both conceptual and mathematical frameworks.}

\smallskip
\smallskip

\noindent {\it  {\sc Empirical observation by subjects (or agents):} The here and now conscious act of observation of a subject (or agent).}

\smallskip
\smallskip

The theory of electromagnetism comprised through Maxwell's equations and the concepts of charge, field, electricity, magnetism, etc., can be only learned in a classroom. As a student in physics one does not go around looking for fields without understanding before what the notion of `field' really means; how it relates to other concepts of the theory or how to mathematically compute the evolution of a specific state of affairs. The understanding of physical notions is always relational, in the sense that a notion always refers to other other notions, like a net in which each node is supported by its neighbor. Only once you grasp the theory of electromagnetism as a whole, it makes sense to go into the lab and try to measure an electromagnetic field. The observation of a field can be only discussed as a complex interplay between theoretical knowledge ---constituted by mathematical equations and concepts--- and what actually happens {\it hic et nunc}. Theories do not come with a user's manual which explains how to measure the `things' the theory talks about. Physics simply does not work like that. Theories do not begin with observations, they end with observations as corroborations of a theoretical representation. It is only once you have a theory that you can understand what is observed. Theoretical measurements can be only performed by subjects (or agents) who understand the theory and the technical devices used in the lab in order to reproduce {\it hic et nunc} an already represented theoretical experience. A physical measurement implies thus the most difficult balance between theory and observation. Theoretical experience requires conceptual and mathematical frameworks, but theories are nothing without an adequate contact to empirical observation. This is what physical (or theoretical) measurement is all about.  

\smallskip
\smallskip

\noindent {\it  {\sc Physical (or Theoretical) Measurement:} The point of contact between the objective theoretical representation of a specific situation and the {\it hic et nunc} conscious act of observation performed by a subject (or agent) who knows the theory and interprets both the experiment and the phenomena accordingly.}

\smallskip
\smallskip

At this point it is also important to clearly distinguish between a theoretical measurement and an operational procedure which can also end up in the prediction of an observation. Like theories, some mathematical models are also capable of predicting specific observations. However, in the case of models there is in general no conceptual-formal unity, and consequently, no consistency achieved by the representation. A very good example of this is Bohr's quantum model of the atom which is a set of ``magical'' rules which allow to compute the spectral lines of the Helium atom. Bohr himself accepted in many discussions with Heisenberg, Pauli and even Schr\"odinger that his model was not a theory; it was not only inconsistent, it simply did not provide a tenable representation of how to think about the atom. It was all these recognized problems that led the efforts of the time to develop a theory with a closed mathematical formalism that would allow to explain and understand the observed phenomena. An operational observation is not linked to a consistent formal conceptual representation. Unlike physical theories which describe `states of affairs' through formal and conceptual moments of unity provided in invariant and objective terms, algorithmic models are only capable of making reference to measurement outcomes, like `clicks' in detectors and `spots' in photographic plates which refer to nothing beyond themselves. 

\smallskip
\smallskip

\noindent {\it  {\sc Operational Observation:} The observation of a measurement result, like a `click' in a detector or a `spot' in a photographic plate, which is predicted in operational terms through a an algorithmic model with no formal nor conceptual unity.}

\smallskip
\smallskip

\noindent An operational observation does not require the reference to a physical concept that captures, as a moment of unity, the phenomena in question. In fact, the reference to `clicks' in detectors in the context of QM, exposes the complete lack of conceptual representation.\footnote{It is not true that `clicks' predicted by the quantum formalism can be regarded ---as Bohr and positivists did--- as being ``classical''. It is Boole-Bell inequalities which have proven that such `clicks' cannot be considered within any classical theory which represents reality as an actual state of affairs (for a detailed discussion see \cite{deRondeMassri18}). The predictions implied by the theory of quanta simply cannot be regarded as arising from the classical presuppositions implied by the theories of Newton and Maxwell. In short, Boole-Bell inequalities must be understood as providing the very conditions of possible classical experience \cite{Pitowsky94}. Aspect' operational measurements have proven that the `clicks' arising in an EPR type experiment cannot be regarded as `classical clicks' (see \cite{Aspect81}). It is interesting to notice that the Bohrian reference to `clicks' in quantum experiments as being ``classical'' is the kernel point that demonstrates that ---after all--- Bohr was much closer to empiricism than to Kantian philosophy.} This is the reason why QM must be still regarded as a proto-theory which, even though possesses a closed mathematical formalism capable of operational quantitative predictions, is still in need of a conceptual representation capable to explain qualitatively the theoretical experience it talks about.\footnote{This is what Tim Maudlin has characterized as a ``practical recipe'' which in the context of QM allows to make predictions from the mathematical formalism. ``What is presented in the average physics textbook, what students learn and researchers use, turns out not to be a precise physical theory at all. It is rather a very effective and accurate recipe for making certain sorts of predictions. What physics students learn is how to use the recipe.''} This should be regarded as one of the main realist (or physical) problems within QM; i.e., the development of a conceptual representation that provides not only a conceptual {\it moment of unity} to the mathematical formalism, but also an {\it anschaulich} (intuitive) content to the theory.

\section{On What `Measurement' is Not: From Bohr to Neo-Bohrians} 

During the re-foundation of physics in the 20th Century, one of the most influential alterations was produced  by Niels Bohr's redefinition of the notion of `measurement' in the context of QM. This new characterization of the meaning of measurement goes hand in hand with what could be called `Bohr's pendular scheme of argumentation' which consisted on a balanced oscillation between an operational (or instrumentalist) reference to `clicks' in detectors and an atomist (metaphysical) narrative grounded on fictional notions such as quantum particles, quantum jumps, etc., both of them connected via the {\it ad hoc} introduction of a {\it correspondence principle}. This allowed Bohr to create a circular justification that remained constantly in motion without ever reaching any reference nor understanding. After many small battles, it was in 1935, that Bohr was able to popularize his pendular scheme within science when most physicists accepted uncritically his reply to the famous EPR paper \cite[p. 1025]{Bohr35} as the final triumph of QM itself over Einstein's conservative remarks. Without making any fuzz of the many lacunas, ambiguities and even inconsistencies within the paper, Bohr was congratulated and applauded by the physicist community as the new champion. In his reply, the Danish physicist had begun by immediately shifting the focus of analysis from EPR's theoretical definition of physical reality to the discussion about the applicability of classical measurement apparatuses. Once the classical representation was introduced within the discussion with extreme detail, he created a story ---with no relation to the mathematical formalism nor to any operational test--- through the introduction of irrepresentable {\it fictional notions} such as `quantum particles', `quantum jumps', `quantum individuality', etc. In this way, according to Bohr's narrative, the measurement in QM was caused by the interaction of elementary quantum particles which affected through quantum jumps the classical measuring apparatuses in a manner that, due to the quantum of action, could not be described by the theory. As he argued \cite[p. 701]{Bohr35}:  ``The impossibility of a closer analysis of the reactions between the particle and the measuring instrument is indeed no peculiarity of the experimental procedure described, but is rather an essential property of any arrangement suited to the study of the phenomena of the type concerned, where we have to do with a feature of [quantum] individuality completely foreign to classical physics.'' Bohr had already applied this ``solution'' in a discussion with Schr\"odinger about the existence of ``quantum jumps'' within the atom. During a meeting in Copenhagen in 1926 under the attentive gaze of Heisenberg, Schr\"odinger  presented several arguments exposing not only the lack of explanation but also the contradictions reached when introducing these a-causal jumps. The lack of theoretical support allowed Schr\"odinger to conclude that ``the whole idea of quantum jumps is sheer fantasy.'' \cite[p. 73]{Heis71} But while the Austrian physicist was not willing to accept the complete lack of theoretical representation within the atom ---a critical position shared only by Einstein---, Bohr was ready to blame the lack of a consistent representation to the theory of quanta itself. As he explained to Schr\"odinger:
\begin{quotation}
\noindent {\small ``What you say is absolutely correct. But it does not prove that there are no quantum jumps. It only proves that we cannot imagine them, that the representational concepts with which we describe events in daily life and experiments in classical physics are inadequate when it comes to describing quantum jumps. Nor should we be surprised to find it so, seeing that the processes involved are not the objects of direct experience.'' \cite[p. 74]{Heis71}} 
\end{quotation}   
According to Bohr, QM went beyond our classical image of the world, and thus ---he argued--- it was no surprise that our (classical) concepts were incapable of explaining what was really going on in the quantum domain. QM referred to a microscopic realm that simply could not be represented. There was nothing to be done but accept the physical and technical limitations we had finally reached within science. According to Bohr, these epistemological restrictions were ontologically imposed by Nature herself. In this way, Bohr turned his own incapacity to develop a consistent representation of QM into a proof of the theory's own difficulties and limits.\footnote{Recognizing the danger of Bohr's scheme Karl Popper \cite{Popper63} strongly criticized Bohr's complementarity solution. Popper hoped that physicists would recognize the untenability of Bohr's proposal: ``I trust that physicists will soon come to realize that the principle of complementarity is {\it ad hoc}, and (what is more important) that its only function is to avoid criticism and to prevent the discussion of physical interpretations; though criticism and discussion are urgently needed for reforming any theory. They will then no longer believe that instrumentalism is forced upon them by the structure of contemporary physical theory.'' Unfortunately, Popper's expectations were not fulfilled. Some decades after Deustch \cite[pp. 309-310]{Deutsch04} characterized the effect of Bohr's scheme: ``For decades, various versions of all that were taught as fact ---vagueness, anthropocentrism, instrumentalism and all--- in university physics courses. Few physicists claimed to understand it. None did, and so students' questions were met with such nonsense as `If you think you've understood quantum mechanics then you don't.' Inconsistency was defended as `complementarity' or `duality'; parochialism was hailed as philosophical sophistication. Thus the theory claimed to stand outside the jurisdiction of normal (i.e. all) modes of criticism ---a hallmark of bad philosophy.''} As a leader of the community, Bohr \cite[p. 7]{WZ} forbid physicists to go beyond the limits imposed by classical concepts. No one was allowed to look for a different solution. As he warned everyone: ``it would be a misconception to believe that the difficulties of the atomic theory may be evaded by eventually replacing the concepts of classical physics by new conceptual forms.'' Of course, as we have just seen, Bohr did not respect his own dictum and repeatedly introduced many non-classical concepts (such as quantum particles, quantum jumps, quantum waves, etc.) which allowed him to create illusions which, in turn, also allowed him to avoid explanations. These newly introduced notions didn't really mean anything, they were not related to the mathematical formalism nor could be operationally tested. However, they did play an essential role within Bohr's scheme, namely, to stop difficult questions. In this respect, David Deutsch, one of the few contemporary physicists who has criticized in depth Bohr's scheme of thought, has remarked the following: 
\begin{quotation}
\noindent {\small ``Let me define `bad philosophy' as philosophy that is not merely false, but actively prevents the growth of other knowledge. In this case, instrumentalism was acting to prevent the explanations in Schr\"odinger's and Heisenberg's theories from being improved or elaborated or unified. The physicist Niels Bohr (another of the pioneers of quantum theory) then developed an `interpretation' of the theory which later became known as the `Copenhagen interpretation'. It said that quantum theory, including the rule of thumb, was a complete description of reality. Bohr excused the various contradictions and gaps by using a combination of instrumentalism and studied ambiguity. He denied the `possibility of speaking of phenomena as existing objectively'  ---but said that only the outcomes of observations should count as phenomena. He also said that, although observation has no access to `the real essence of phenomena', it does reveal relationships between them, and that, in addition, quantum theory blurs the distinction between observer and observed. As for what would happen if one observer performed a quantum-level observation on another, he avoided the issue [...]'' \cite[p. 308]{Deutsch04}} 
\end{quotation}    
The Bohrian technique of argumentation consisted in discussing on two disconnected parallel levels jumping back and forth from one to the other, avoiding in this way questions that could not be answered. The methodology is quite simple, if a question cannot be answered in one level, simply shift to the other. Moving from our ``common sense'' manifest image of the world to fictitious stories about a quantum realm that could not be represented, Bohr was able to create the illusion of understanding. But in the end, his circular analysis begun with classically described experimental arrangements and ended up in the prediction of `clicks' in detectors. In this respect, as remarked by Deutsch, Bohr's scheme might reminds us of conjuring tricks rather than scientific explanation. 
\begin{quotation}
\noindent {\small ``Some people may enjoy conjuring tricks without ever wanting to
know how they work. Similarly, during the twentieth century, most
philosophers, and many scientists, took the view that science is incapable
of discovering anything about reality. Starting from empiricism, they
drew the inevitable conclusion (which would nevertheless have horrified
the early empiricists) that science cannot validly do more than predict
the outcomes of observations, and that it should never purport to
describe the reality that brings those outcomes about. This is known as
instrumentalism. It denies that what I have been calling `explanation'
can exist at all. It is still very influential. In some fields (such as statistical
analysis) the very word `explanation' has come to mean prediction, so
that a mathematical formula is said to `explain' a set of experimental
data. By `reality' is meant merely the observed data that the formula is
supposed to approximate. That leaves no term for assertions about
reality itself, except perhaps `useful fiction'.'' \cite[p. 15]{Deutsch04}} 
\end{quotation}    
In this way, theoretical explanation became confused with the {\it ad hoc} postulation of rules and  inconsistent interpretations became confused with theoretical explanation. Bohr created a fictional quantum narrative which allowed physicists to believe in something they did not understand. The power of Bohr's pendular scheme was that it allowed a dual discourse. While the realist believer could claim that quantum particles truly existed and were responsible for the macroscopic world we observe, the anti-realist was allowed to maintain their sceptic position that QM only made reference to the prediction of `clicks' in detectors or `spots' in photographic plates. Both assertions co-existed in Bohr's inconsistent rhetorics. The replacement of theoretical (formal-conceptual) representation by {\it ad hoc} rules and fictional stories meant also the abandonment of the methodology and goals implied by physical research (section 4). Bohr's program, in line with the positivist {\it Zeitgeist} of the 20th century, was then ready to take a step further and redefine the meaning of `physics' itself: 
\begin{quotation}
\noindent {\small ``Physics is to be regarded not so much as the study of something a priori given, but rather as {\small{\it the development of methods of ordering and surveying human experience.}} In this respect our task must be to account for such experience in a manner independent of individual subjective judgement and therefor {\small{\it objective\footnote{`Objective' for Bohr means in fact `intersubjective'. See section 6.3.} in the sense that it can be unambiguously communicated in ordinary human language.}}'' \cite{Bohr60} (emphasis added)} 
\end{quotation}
This shift ---which was already part of the positivist scheme of thought--- turned physics away from {\it physis} and closer to the (empirical) subject's ``common sense'' observations and measurements. Since the concept of measurement had always made reference to physical reality, the redefinition of a new `physics' detached from {\it physis} implied also the need to redefine the notion of `measurement' accordingly. Making use of his pendular rhetorics, Bohr provided two confronting definitions of `measurement'. On the one hand, he \cite[p. 209]{Bohr49} related measurement to the intersubjective communication of observations by (empirical) subjects arguing that ``by the word `experiment' we refer to a situation where we can tell others what we have done and what we have learned and that, therefore, the account of the experimental arrangement and of the results of the observations must be expressed in unambiguous language with suitable application of the terminology of classical physics.'' 

\smallskip
\smallskip

\noindent {\it  {\sc Conscious Observation-Communication Measurement:} The here and now conscious act of observation of a subject (or agent) who is capable to communicate what she observed to other subjects (or agents).}

\smallskip
\smallskip

\noindent On the other hand, Bohr also made reference to an understanding of `measurement' in terms of `interacting systems'. Shifting to a metaphysical level of description, it was now assumed that a measurement was essentially a process in which a `click' in a detector or the `imprint' in a photographic plate was produced due to the interaction between a quantum particle and a classical apparatus: ``In fact to measure the position of one of the particles can mean nothing else than to establish a correlation between its behavior and some instrument rigidly fixed to the support which defines the space frame of reference.'' Once again, the existence of this process was constructed in a fictional manner without providing any consistent link to the mathematical formalism of the theory. 
\smallskip
\smallskip

\noindent {\it  {\sc Systems Interaction-Correlation Measurement:} The interaction between a system and an apparatus produced by an incontrollable interaction (due to the quantum of action) which allows a correlation between them. This process, independent of subjects, ends up in the imprint of a `spot' in a photographic plate or the sound of a `click' in a detector.}

\smallskip
\smallskip

\noindent Adding to the confusion, in his pendular fashion, Bohr argued: 
\begin{quotation}
\noindent {\small ``As we have seen, any observation necessitates an interference with the course of the phenomena, which is of such a nature that it deprives us of the foundation underlying the causal mode of description. The limit, which nature herself has thus imposed upon us, of the possibility of speaking about phenomena as existing objectively finds its expression, as far as we can judge, just in the formulation of quantum mechanics. The discovery of the quantum of action shows us, in fact, not only the natural limitation of classical physics, but, by throwing a new light upon the old philosophical problem of the objective existence of phenomena independently of our observations, confronts us with a situation hitherto unknown in natural science.'' 
\cite[p. 115]{Bohr34}}
\end{quotation}

Quite irrespectively of the internal contradictions, inconsistencies and ambiguities, Bohr's pendular scheme has continued to play an essential role within the contemporary debates about QM. Today, there are many interpretations exposing the still present impact of Bohr's ideas, arguments and reasoning within the specialized foundational and philosophical literature of the 21st Centrury. In the following, we are interested in discussing three different neo-Bohrian approaches which address explicitly the question of measurement in QM: Zurek's model of decoherence, modal interpretations and QBism. 

\subsection{A Neo-Bohrian ``Technical Solution'': Decoherence}

From a positivist viewpoint which tends to focus in the analysis of formal (mathematical and logical) models which are able to account for empirical observations, one of the main problems in the writings of Bohr is the complete lack of any reference to the orthodox mathematical formalism of QM. Bohr evaded such an analysis and shifted his attention ---and that of the community--- to the analysis of the already known classical measurement situations. He justified himself by explaining that after all \cite[p. 7]{WZ}: ``[...] the unambiguous interpretation  of any measurement must be essentially framed in terms of classical physical theories, and we may say that in this sense the language of Newton and Maxwell will remain the language of physicists for all time.'' At the beginning of the 1970s, attempting to fill this formal void, Dieter Zhe \cite{Zhe70} discussed the meaning of measurement in QM in a paper that would pioneer what would be later on known as decoherence. Decoherence was conceived by Zeh as a model that would finally bridge the gap between QM and our classical macroscopic world; explaining how weird microscopic quantum superpositions could end up transforming themselves into tables and chairs. During the 1980s, Wojciech Zurek popularized these ideas relating them more explicitly to Bohr's understanding(s) of measurement and the quantum to classical limit \cite{Zurek81, Zurek82}. There is an essential circularity in the proposal since the application of particle metaphysics to the quantum formalism is what needs to be explained, instead of presupposed. One can easily understand the way in which ``small particles'' are able to constitute ``big objects'', what is difficult is to explain is the way in which ``small particles'' can be represented  in terms of quantum superpositions or how entangled states can become a table.

Using an analogy with classical thermodynamics with no obvious link to the mathematical formalism of QM a new fictional notion was added to the scheme of decoherence, namely, that of {\it quantum harmonic oscillator}. By considering an infinite sum of such ``harmonic oscillators'' the ``coherence'' of quantum particles was expected to decrease. Adding to the confusion, Everett's relative state formulation ---closely connected to Bohr's ideas on contextuality--- was another essential condiment of the original scheme. However, regardless of the efforts to create a narrative, according to the formalism decoherence never actually took place. It was found that a sum of quantum states retain their entanglement independently of how many of them are considered. Even an infinite sum of quantum states do not de-cohere. At this kernel point, in a complete Bohrian fashion, it was argued that what was in fact needed was the addition of a (classical) ``continuum bath'' also called ``environment''. After all ---it was argued---, what surrounds the quantum microscopic realm is our classical macroscopic one. This shift from a (quantum) {\it discrete} representation ---consequence of Planck's quantum postulate--- to a (classical) {\it continuous} representation might seem to the attentive reader like a desperate attempt to impose a solution ---instead of finding it. Indeed, an infinite numerable sum of Hamiltonians of elementary harmonic oscillators with natural frequencies\footnote{It should be remarked that the notion of harmonic oscillator has a clear meaning within classical physics; however, its extension to quantum mechanics is far from evident. If QM does not describe `particles' nor `waves', what is then oscillating?} is not the same as an integral of the Hamiltonians of a continuum of oscillators with real frequencies. It is this mathematical addition of the continuous which hides the {\it ad hoc} imposition a classical representation. That which needed to be physically explained in conceptual and formal terms ---i.e., the appearance of the classical from the quantum formalism--- was simply presupposed as a natural addition. The introduction of this ``continuum bath'' of harmonic oscillators was then justified through the idea that an ``open system'' (i.e., a great number of interacting systems) is ``more real'', or ``less idealized'', than a ``closed system'' (i.e., a completely isolated system). The argument points to the seemingly obvious fact that what actually happens in our world is that systems are always in interaction with other systems. Due to technical reasons it might be impossible to ``close a system'' completely and that is the reason why quantum states become classical. After going beyond representation and treating systems as `things' in the real world, the argument also makes use of the Bohrian-positivist presupposition according to which an `object', in order to bear existence, must be observed by someone or something. As Bohr stressed repeatedly, a quantum system cannot be described independently of its context of existence, it must be always related to a classical apparatus or ---in the case of decoherence--- to the environment. The inconsistency present within these ideas is very extreme. 

As we have argued above, a theoretical representation characterizes the existents it talks about and the interactions between them are just part of the representation; a representation cannot be considered as ``open'' or ``closed''. The representation of a ``single particle'', is as abstract as that of ``many particles''. In fact, one cannot think of ``many particles'' without presupposing the representation of a ``single particle''. Thermodynamics is a generalization of classical mechanics, but it would be ridiculous to claim it is ``more real'' than classical mechanics since it makes only sense by presupposing it. It is simply ridiculous to claim that if we consider many particles the representation becomes ``more real'' than if we consider only one particle ---since the first presupposes the latter. Furthermore, there exists according to QM an intrinsic discreteness due to the quantum postulate inherent within the mathematical formalism which makes the description of a single particle untenable. In fact, Max Planck created QM by replacing an integer of {\it continuous} energy by a sum of quantum packages of {\it discrete} energies. That is the whole point about the {\it quantum}, namely, that it is {\it discrete}. Making use of the ``common sense'' atomist idea that it is impossible to ``close a system'' completely, the discrete sum of superpositions is blamed for being part of an ``inaccurate'' representation of reality. As if the idea of ``open system'' ---which is nothing but many interacting (closed) systems--- was not part of the same representation, the addition of the continuum becomes then a necessary condition for a more accurate representation of things {\it as they really are}. Decoherence not only shows the complete lack of comprehension regarding the scope of theoretical representation and measurement, it also exposes the inconsistency of the program itself for if quantum superpositions are ``less real'' than the environment ---which is ``closer to reality''--- why should we even bother in finding a limit? 

In addition to these unjustifiable formal jumps and {\it ad hoc} maneuvers in the conceptual and formal descriptions, there are many other technical aspects which also expose the failure of program. The fact that the diagonalization is not complete, since ``very small'' is obviously not ``equal to zero'',\footnote{Notice that within an epistemological account, ``very small'' might be considered as superfluous when compared to ``very big''; however, this is clearly not  the case from an ontological account. From an ontological perspective there is no essential difference between ``very big'' and ``very small'', they both have exactly the same importance.} the fact that the diagonalization can recompose itself into un-diagonalized mixtures if enough time is considered \cite{Recoherence, CormikPaz08} and the fact that the principle turns (non-diagonal) improper mixtures into (``approximately'' diagonal) improper mixtures which still cannot be interpreted in terms of ignorance are just a few of the many failures of the program of decoherence.\footnote{The late recognition of this fact by Zurek has lead him to venture into many worlds interpretation in which case there are also serious inconsistencies threatening the project \cite{DawinThebault15}. Ruth Kastner has even pointed out quite clearly why ---even if these many points would be left aside--- the main reasoning of the decoherence program is circular \cite{Kastner14}.} After many criticisms within the specialized literature, and just to make things even less clear,  Zurek  \cite[p. 22]{Zurek02} ---in a truly Bohrian spirit--- decided to introduce also a subjective account of measurement by arguing that: ``Quantum state vectors can be real, but only when the superposition principle ---a cornerstone of quantum behavior--- is `turned off' by einselection. Yet einselection is caused by the transfer of information about selected observables. Hence, the ontological features of the state vectors ---objective existence of the einselected states--- is acquired through the epistemological `information transfer'.''

It was due to the insistent critical analysis coming mainly from philosophers of QM, that the failure of the decoherence program to explain the quantum to classical limit and the measurement problem had to be recognized by its supporters.\footnote{Regardless of this, as remarked by Guido Bacciagaluppi \cite{Bacc12}: ``[some physicists and philosophers] still believe decoherence would provide a solution to the measurement problem of quantum mechanics. As pointed out by many authors, however (e.g. Adler 2003; Zeh 1995, pp. 14-15), this claim is not tenable. [...] Unfortunately, naive claims of the kind that decoherence gives a complete answer to the measurement problem are still somewhat part of the `folklore' of decoherence, and deservedly attract the wrath of physicists (e.g. Pearle 1997) and philosophers (e.g. Bub 1997, Chap. 8) alike.'' } But in an amazing rhetorical move ---which reminds us Bohr's method---, making use of the fact that some operational results and models had been already constructed, decoherent theorists argued that even though decoherence did not provide a theoretical explanation of the quantum to classical limit nor of the measurement problem, it did provide a solution ``For All Practical Purposes'' (a ``FAPP solution''). By creating a new type of ``solution'', the process of decoherence justified its own existence and became part ---together with ``quantum jumps'' and ``quantum particles''--- of the contemporary postmodern narrative of quantum physics. Today, the widespread acceptance of the decoherence shows the influence of Bohr's fictional-instrumentalist legacy. The creation of a new science which does not need to provide a consistent theoretical account of what it talks about and allows to justify itself through the application of operational models supported by a fragmented inconsistent compound of fictions and illusions.

\subsection{A Neo-Bohrian ``Formal Solution'': Modal Interpretations}

Almost concomitant with the creation of decoherence, at the beginning of the 1980s Bas van Fraassen, one of today's most influential contemporary empiricists, proposed another Neo-Bohrian interpretation of QM. Van Fraassen's main idea was to introduce modal logics in order to provide a consistent account of the theory in empiricist terms \cite[pp. 202-203]{VF80} which meant for him: ``to withhold belief in anything that goes beyond the actual, observable phenomena, and to recognize no objective modality in nature. To develop an empiricist account of science is to depict it as involving a search for truth only about the empirical world, about what is actual and observable.'' Following Bohr's ideas regarding the purely algorithmic understanding of the quantum wave function, van Fraassen \cite[p. 288]{VF91} argued that:  ``[the emergence of a result is] {\it as if the Projection Postulate were correct}. For at the end of a measurement of ${\bf A}$ on system $X$, it is indeed true that  ${\bf A}$ has the actual value which is the measurement outcome. But, of course, the Projection Postulate is not really correct: there has been a transition from possible to actual value, so what it entailed about values of observables is correct, but that is all. There has been no acausal state transition.'' Since the {\it possible} was addressed in the terms of modal logics, van Fraassen's interpretation became known as the ``modal interpretation''. Following the same line of reasoning, Dennis Dieks has recently tried to explain why the quantum measurement problem was never explicitly considered by the Danish physicist:\footnote{Regarding the measurement problem, Petersen \cite[p. 249]{OsnaghiFreitasFreire09} in his letter of 1957 to Hugh Everett argued in the same line as Dieks: ``There can on [Bohr's] view be no special observational problem in quantum mechanics in accordance with the fact that the very idea of observation belongs to the frame of classical concepts. The aim of [Bohr's] analysis is only to make explicit what the formalism implies about the application of the elementary physical concepts. The requirement that these concepts are indispensable for an unambiguous account of the observations is met without further assumptions [...].''}
\begin{quotation}
\noindent {\small ``[...] measuring devices, like all macroscopic objects around us, can and must be described classically. It is an immediate consequence of this that measurements necessarily have only one single outcome. Pointers can only have one position at a time, a light flashes or does not flash, and so on ---this is all inherent in the uniqueness of the classical description. Because of this, Bohr's interpretation does not face the ``measurement problem'' in the form in which it is often posed in the foundational literature, namely as the problem of how to explain ---in the face of the presence of superpositions in the mathematical formalism--- that there is only one outcome realized each time we run an experiment. For Bohr this is not something to be explained, but rather something that is given and has to be assumed to start with. It is a primitive datum, in the same sense that the applicability of classical language to our everyday world is a brute fact to which the interpretation of quantum mechanics necessarily has to conform. An interpretation that would predict that pointers can have more than one position, that a cat can be both dead and alive, etc., would be a non-starter from Bohr's point of view. So the measurement problem in its usual form does not exist; it is dissolved.'' \cite[p. 24]{Dieks16}}
\end{quotation}
This dissolution of the measurement problem simply re-states the naive empiricist standpoint present in the empirical-positivist and Bohrian understanding of theories in general, and of QM in particular. That's fair. But going beyond the empiricist reference to actualities, Dieks himself advanced in the late 1980s a ``realist'' version of the modal interpretation in which quantum particles were explicitly addressed. Rather than making reference to measurement outcomes, QM should be better understood ``in terms of properties possessed by physical systems, independently of consciousness and measurements (in the sense of human interventions)'' \cite{Dieks07}. Following Bohr's ideas on contextuality and Simon Kochen's proposal \cite{Kochen85}, Dieks took the Schmidt (bi-orthogonal) decomposition and preferred basis as a necessary standpoint for interpreting QM.\footnote{It is interesting to notice that Carl Friedrich von Weizs\"acker and Theodor G\"ornitz \cite[p. 357]{GornitzWeizsacker} referred specifically to Kochen's proposal in a paper entitled ``Remarks on S. Kochen's Interpretation of Quantum Mechanics''. In this paper they state: ``We consider it is an illuminating clarification of the mathematical structure of the theory, especially apt to describe the measuring process. We would, however feel that it means not an alternative but a continuation to the Copenhagen interpretation (Bohr and, to some extent, Heisenberg).''} 
\begin{theo}\label{Schmidt}
Given a state $|\Psi_{\alpha\beta}\rangle$ in $\cal H = \cal
H_{\alpha}\otimes \cal H_{\beta}$. The Schmidt theorem assures there
always exist orthonormal bases for $\cal H_{\alpha}$ and $\cal
H_{\beta}$, $\{|a_{i}\rangle\}$ and $\{|b_{j}\rangle\}$ such that
$|\Psi_{\alpha\beta}\rangle$ can be written as

\begin{center}$|\Psi_{\alpha\beta}\rangle = \sum c_{j}|a_{j}\rangle
\otimes |b_{j}\rangle$.\end{center}

\noindent The different values in $\{|c_{j}|^{2}\}$ represent the
spectrum of the state. Every $\lambda_{j}$ represents a projection
in $\cal H_{\alpha}$ and a projection in $\cal H_{\beta}$ defined as
$P_{\alpha}(\lambda_{j}) = \sum |a_{j}\rangle \langle a_{j}|$ and
$P_{\beta}(\lambda_{j}) = \sum |b_{j}\rangle \langle b_{j}|$,
respectively. Furthermore, if the $\{|c_{j}|^{2}\}$ are non
degenerate, there is a one-to-one correlation between the
projections $P_{\alpha} = \sum |a_{j}\rangle \langle a_{j}|$ and
$P_{\beta} = \sum |b_{j}\rangle \langle b_{j}|$ pertaining to
subsystems $\cal H_{\alpha}$ and $\cal H_{\beta}$ given by each
value of the spectrum.\qed
\end{theo}

\noindent If we assume non-degeneracy the modal interpretation based on the Schmidt decomposition establishes a one-to-one correlation between the reduced states of `system' and `apparatus'. As noted by Kochen \cite[p. 152]{Kochen85}: ``Every interaction gives rise to a unique correlation between certain canonically defined properties of the two interacting systems. These properties form a Boolean algebra and so obey the laws of classical logic.'' The bi-orthogonal decomposition provides in this way a one-to-one correlation between apparatus and (quantum) system according to the following interpretation: {\it The system $\alpha$ possibly possesses one of the properties $\{|a_{j}\rangle \langle a_{j}|\}$, and the actual possessed property $|a_{k}\rangle \langle a_{k}|$ is determined by the device possessing the reading $|b_{k}\rangle \langle b_{k}|$.}\footnote{As Bacciagaluppi \cite{Bacciagaluppi96} points out with respect to Kochen's interpretation: ``It seems that he conceives [the states in the Schmidt decomposition] rather as states that are relative to each other. It seems that he espouses an Everettian view that systems have states only relative to each other but that he considers the ascription of relative states (in Everett sense) only in the symmetrical situation in which not only the $|a_{j}\rangle$ are relative to the $|a_{i}\rangle$, but at the same time the $|a_{j}\rangle$ are relative to the $|a_{i}\rangle$.''} However, there is an essential drawback. Tracing over the degrees of freedom of the system, one obtains an \emph{improper mixture} which cannot be interpreted in terms of ignorance \cite{D'Espagnat76}. Consequently, the path from the possible to the actual cannot be regarded as making reference to an underlying preexistent system with definite valued properties. At this crucial point, in a truly Bohrian spirit, Dieks returned to the safety of an instrumentalist account of the theory which ---leaving `quantum systems' behind--- goes back to the possibility of predicting measurement outcomes.\footnote{In order to do so, while van Fraassen's distinguishes between {\it dynamical states} and {\it value states} \cite{VF91}, Dieks and Vermaas do exactly the same by distinguishing between {\it mathematical states} and {\it physical states} \cite{VermaasDieks95}.} The path from the mathematical formalism to the actual observation was then postulated by Dieks in terms of an {\it ad hoc} algorithmic rule completely equivalent to the projection postulate: 
\begin{quotation}
\noindent {\small ``I now propose the following interpretational rule: as soon as there is a unique decomposition of the form [$|\Psi_{\alpha\beta}\rangle = \sum c_{j}|\phi_{j}\rangle \otimes |R_{j}\rangle$], the partial system represented by the $|\phi_{k}\rangle$, taken by itself, can be described as possessing one of the values of the physical quantity corresponding to the set ${|\phi_{k}\rangle}$, with probability $|c_{k}|^{2}$. 

This rule is intended to have the following important consequence. Experimental data that pertain only to the object system, and that say it possesses the property associated with, e.g., $|\phi_{1}\rangle$, not only count as support for the theoretical description $|\phi_{1}\rangle|R_{1}\rangle$ but also as empirical support for the theoretical description (2).'' \cite[p. 39]{Dieks88b}}
\end{quotation}
Instead of using the mathematical formalism in order to describe systems with definite properties and the way they end up producing `clicks', the formalism was suddenly applied in an instrumentalist fashion in order to predict the actual observation of measurement outcomes. Dieks \cite[p. 182]{Dieks88a} argued that ``[...] there is no need for the projection postulate. On the theoretical level the full superposition of states is always maintained, and the time evolution is unitary. One could say that the `projection' has been shifted from the level of the theoretical formalism to the semantics: it is only the empirical interpretation of the superposition that the component terms sometimes, and to some extent, receive an independent status.'' The {\it projection postulate} was renamed as an ``interpretational rule'' which was accepted at the level of operational prediction but rejected at the level of the realist interpretation. Suddenly, the reference to `systems' and `properties' disappeared and Dieks  \cite[p. 177]{Dieks88a} went back to an analysis about measurement outcomes: ``[...] an irreducible statistical theory only speaks about possible outcomes, not about the actual one; this predicts only probability distributions of all outcomes, and says nothing about the result which really will be realized in a single case. In brief, such a theory is not about what is real and actual but only about what could be the case.'' As it becomes clear, this instrumentalist interpretation of quantum modalities makes no contact with an underlying realist interpretation about `systems'. Dieks ---following the Bohrian scheme--- goes back and forth between a realist interpretation of the mathematical formalism which supposedly describes quantum systems (independently of measurement outcomes) and an empiricist pragmatic interpretation of the same formalism used as a black box in order to predict actual measurement outcomes (see \cite{Dieks10}). The failure of the project extends itself to the {\it preferred} Schmidt basis which becomes only essential in creating a new formal narrative about quantum systems that interact with classical apparatuses.  
\begin{quotation}
\noindent {\small ``[...] there should be a one-to-one correlation between the definite properties of a system and the definite properties of its environment [...] can be seen as a way to generalize (and make rigorous) a significant part of Bohr's interpretation of quantum mechanics. According to Bohr the applicability of concepts depends on the type of macroscopic measuring device that is present; given a `phenomenon' there is a one- to-one correspondence between the properties of the measuring device and those of the object system. Our second requirement implements this idea also in situations in which there is no macroscopic measuring device, but only a correlation with the (possibly microscopic) environment.

The idea that there is a correspondence between properties of a system and those of its environment is also physically motivated by other approaches to the interpretation of quantum mechanics, especially the decoherence approach (the `monitoring' of a system by its environment, see Ref. [5] and references therein).'' 
\cite[p. 368]{Dieks95}}
\end{quotation}

Independently of the initial lacunas and ambiguities, during the 1990s Dieks' modal realistic reference to systems was confronted to the mathematical formalism of the theory itself. It is interesting to notice that this type of metaphysical analysis and research had not been explicitly performed in QM since Einstein's and Schr\"odinger's critical work during the 1930s. As a deeply provocative result, many no-go theorems specifically designed for modal interpretations begun to expose the already known difficulties to link the orthodox mathematical formalism with the notion of `system' ---i.e., a physical entity composed of definite valued properties.\footnote{We refer the interested reader to the detailed analysis presented in \cite{deRonde11}. For an analysis of Kochen-Specker contextuality and its implications for the interpretation of projection operators see \cite{deRonde19}.} Like in the case of Bohr, Dieks was then forced to assume an even more extreme form of relativism ---already implicit in the self imposed restriction to the Schmidt preferred basis. Dieks perspectivalist proposal co-authored with Gyula Bene, added a further non-invariant reference through the choice of a new factorization of the total Hilbert space \cite{BeneDieks02}. As in the case of decoherence, the essential difficulty of the modal interpretation is that even though there might seem to exist an account of correlated systems, there is no independent representation of the constituents. So even though the `interpretation' attempts to evade an account of `single systems' ---through the reference to the bi-orthogonal decomposition and factorizations--- it nevertheless talks about composite ones creating the illusion of a reference which simply does not exist. The realist version of the modal interpretation fails to explain the basics. What is a `quantum system'? What is a `classical apparatus'? How does a `quantum system' interact with a `classical system' or how is a single `click' generated by a particle? A realist interpretation should be able to answer these questions instead of providing a instrumentalist reply which makes only reference to the prediction of outcomes.

\subsection{A Neo-Bohrian ``Subjectivist Solution'': QBism}

Bohr's understanding of the mathematical formalism of QM as making exclusive reference to measurement outcomes had as a consequence a radical shift from the {\it objective} character of theoretical representation ---in terms of conceptual moments of unity--- to the {\it intersubjective} communication of individual observations between empirical subjects. In this way, the experiences acquired by different agents, were detached from a common {\it objective} reference and representation.\footnote{In his book, \cite{D'Espagnat06}, D'Espagnat clearly distinguishes between {\it objective statements} and Bohr's {\it intersubjective statements},  which he calls: {\it weakly objective statements}.} This also implied a silent replacement from the notion of {\it object} ---categorically constituted through the general principles of existence, non-contradiction and identity---; by that of {\it event} (e.g., `clicks' and `spots') ---which has no categorical nor conceptual constitution.\footnote{While the object acts as a moment of unity that is able to account for the multiplicity of experience, the `click' makes only reference to a fragmented experience with no internal unity.} Bohr shifted from objectivity to intersubjectivity, but he was not willing to abandon the term `objective'. So instead of calling things by their name he simply renamed `intersubjective statements' and called them `objective statements'. Stressing the claim that his interpretation of QM was as objective as in classical physics he \cite[p. 98]{D'Espagnat06} argued that: ``The description of atomic phenomena has [...] a perfectly objective character, in the sense that no explicit reference is made to any individual observer and that therefore... no ambiguity is involved in the communication of observation.'' Bernard D'Espagnat explains this quotation in the following manner: ``That Bohr identified objectivity with intersubjectivity is a fact that the quotation above makes crystal clear. In view of this, one cannot fail to be surprised by the large number of his commentators, including competent ones, who merely half-agree on this, and only with ambiguous words. It seem they could not resign themselves to the ominous fact that Bohr was not a realist.'' 

With the arrival of the new millennium and in tune with the anti-realist postmodern {\it Zeitgeist} of the 20th Century ---which had maliciously confused `realism' either a correspondence between theory  reality itself' or the belief of a subject---, Quantum Bayesianism (QBism for short) was developed by a group of researchers as one of the most honest neo-Bohrian approaches to QM.  Following Bohr, QBism \cite[p. 70]{FuchsPeres00} took as a standpoint the unspoken divorce between quantum theory and physical reality: ``[...] quantum theory does not describe physical reality. What it does is provide an algorithm for computing probabilities for the macroscopic events (`detector clicks') that are the consequences of experimental interventions.'' As made explicit by Chris Fuchs \cite{QBism13}: ``QBism agrees with Bohr that the primitive concept of experience is fundamental to an understanding of science.''\footnote{In recent years David Mermin has become also part of the QBist team, publishing several papers which not only support, but also make clear the connection of QBism to the Bohrian interpretation of QM. See: \cite{Mermin04, Mermin14}.} It is in this context that the Bayesian subjectivist interpretation of probability is introduced.
\begin{quotation}
\noindent {\small ``QBism explicitly takes the `subjective' or `judgmental' or `personalist' view of probability, which, though common among contemporary statisticians and economists, is still rare among physicists: probabilities are assigned to an event by an agent and are particular to that agent. The agent's probability assignments express her own personal degrees of belief about the event. The personal character of probability includes cases in which the agent is certain about the event: even probabilities 0 and 1 are measures of an agent's (very strongly held) belief.'' \cite[p. 750]{QBism13}}
\end{quotation}

Regardless of our obvious philosophical distance with respect to QBism, we believe that this purely instrumentalist account of QM avoids the pendular reference imposed by Bohr's rhetorics. At least in a first stage, QBism allows to make explicitly clear the subjectivist standpoint implied by anti-realist approaches.\footnote{Recently, Chris Fuchs has changed his anti-realist claims and presented something called ``participatory realism''.} The consistency of their approach is secured by their explicit denial of any reference to physical reality ---something that not many anti-realists are eager to accept. The QBist understanding of measurement is then exclusively related to a purely conscious act of observation [{\it Op. cit.}, p. 750]: ``A measurement in QBism is more than a procedure in a laboratory. It is any action an agent takes to elicit a set of possible experiences. The measurement outcome is the particular experience of that agent elicited in this way. Given a measurement outcome, the quantum formalism guides the agent in updating her probabilities for subsequent measurements.'' Indeed, as QBist make explicitly clear: ``A measurement does not, as the term unfortunately suggests, reveal a pre-existing state of affairs.'' Measurements are personal, individual and QM is a ``tool'' for the ``user''  ---as Mermin prefers to call the ``agent'' \cite{Mermin14}. Just like a mobile phone or a laptop, QM is a  tool that we subjects use in order to organize our experience.
\begin{quotation}
\noindent {\small ``QBist takes quantum mechanics to be a personal mode of thought ---a very powerful tool that any agent can use to organize her own experience. That each of us can use such a tool to organize our own experience with spectacular success is an extremely important objective fact about the world we live in. But quantum mechanics itself does not deal directly with the objective world; it deals with the experiences of that objective world that belong to whatever particular agent is making use of the quantum theory.'' [{\it Op. cit.}, p. 751]}
\end{quotation}
According to Fuchs, measurement is an {\it irrepresentable} ``act of creation'': 
\begin{quotation}
\noindent {\small ``The research program of Quantum Bayesianism (or QBism) is an approach to quantum theory that hopes to show with mathematical precision that its greatest lesson is the world's plasticity. With every quantum measurement set by an experimenter's free will, the world is shaped just a little as it takes part in a moment of creation. So too it is with every action of every agent everywhere, not just experimentalists in laboratories. Quantum measurement represents those moments of creation that are sought out or noticed.'' \cite[p. 2114]{Fuchs14}}
\end{quotation}
QBism ---following Bohr--- accepts the failure of science to provide a representation of the world we live in.  
\begin{quotation}
\noindent {\small ``[...] a measurement apparatus must be understood as an extension of the agent herself, not something foreign and separate. A quantum measurement device is like a prosthetic hand, and the outcome of a measurement is an unpredictable, undetermined `experience' shared between the agent and external system. Quantum theory, thus, is no mirror image of what the world is, but rather a `user's manual' that any agent can adopt for better navigation in a world suffused with creation: The agent uses it for her little part and participation in this creation.'' \cite[p. 2042]{Fuchs14}}
\end{quotation}
The simple question that rises is the following: if a physical theory like QM does not provide understanding of the world and Nature, what discipline is supposed to do this? And if the stories told by science are not making reference to reality, to particles, fields or other type of entities, should we simply accept that all we observe as subjects is part of an illusion? A fictitious story we created ourselves?

\section{The (Quantum) Measurement Problem in Classical Mechanics}

As we discussed above, the QMP is grounded on one of the main empiricist pillars of positivism according to which a theory must account for what is uncritically observed in the lab. This naive empiricist standpoint according to which there can be a ``direct access'' to the world that surround us by ``simply observing what is going on'' was fantastically addressed ---and ironically criticized--- by the Argentine writer Jorge Luis Borges in a beautiful short story called {\it Funes the memorious} \cite{Borges}.\footnote{The first cornerstone  of positivism which was also partially accepted by Bohr's claim that experience is always ``classical''. Even though Bohr's neo-Kantian approach recognized the fact that classical experience was not a {\it given} but rather a complex constitution provided by the subjects transcendental categories and forms of intuition, the prohibition to consider any other experience placed observations in the same place as positivists: as the fundament of all possible scientific knowledge. This coincidence is what allowed to seal the silent alliance between positivists and neo-Kantians.} Borges recalls his encounter with Ireneo Funes, a young man from Fray Bentos who after having an accident become paralyzed. Since then, Funes' perception and memory became infallible. According to Borges, the least important of his recollections was more minutely precise and more lively than our perception of a physical pleasure or a physical torment. However, as Borges also remarked: ``He was, let us not forget, almost incapable of general, platonic ideas. It was not only difficult for him to understand that the generic term dog embraced so many unlike specimens of differing sizes and different forms; he was disturbed by the fact that a dog at three-fourteen (seen in profile) should have the same name as the dog at three fifteen (seen from the front). [...] Without effort, he had learned English, French, Portuguese, Latin. I suspect, however, that he was not very capable of thought. To think is to forget differences, generalize, make abstractions. In the teeming world of Funes there were only details, almost immediate in their presence.'' Using the story as a {\it Gedankenexperiment} Borges shows why, for a radical empiricist as Funes, there is no reason to believe in the (metaphysical) identity of `the dog at three-fourteen (seen in profile)' and `the dog at three fifteen (seen from the front)'. For Funes, the radical empiricist capable of apprehending experience beyond conceptual presuppositions, there is no `dog'; simply because experience does not contain the (metaphysical) {\it moment of unity} required to make reference to {\it the same} through time. ``Locke, in the seventeenth century, postulated (and rejected) an impossible language in which each individual thing, each stone, each bird and each branch, would have its own name; Funes once projected an analogous language, but discarded it because it seemed too general to him, too ambiguous. In fact, Funes remembered not only every leaf of every tree of every wood, but also every one of the times he had perceived or imagined it.'' It is through Aristotle's metaphysics and the application of the ontological and logical principles of existence, identity and non-contradiction, that science was able to provide a conceptual architectonic which allows us to connect the `the dog at three-fourteen (seen in profile)' and `the dog at three fifteen (seen from the front)' in terms of a {\it sameness}. In general, it is only by presupposing these principles that we can think in terms of `individual systems' ---such as, for example, a `dog'. Borges shows why these principles are not self-evident {\it givens} of experience, and neither is a `dog'. And this is the reason why Borges also suspected that Funes ``was not very capable of thought.'' Funes might be regarded as the true non-naive empiricist, who understands that every observation is just a fragmented conglomerate of sensations with no internal unity. Returning now to our discussion, what is important to understand is that the QMP does not make reference to theoretical measurements, it makes reference to operational observations of `clicks' in detectors and `spots' in photographic plates. Such `clicks' and `spots' play exactly the same role as Funes' pure fragmented observations. Like Funes' static observations of pure sensation a `click' has no reference beyond itself. A `click' is not a {\it moment of unity}, it has no metaphysical constitution, and consequently it escapes theoretical representation right from the start. No wonder, it becomes then impossible to relate it to a physical concept or to the mathematical formalism. 

We are now able to expose the untenability of the so called measurement problem which, in fact, should be called ``the observability problem''. If we assume that observations can be singular and without a conceptual moment of unity, like in the case of a `click' in a detector, then an analogous measurement problem can be derived also in the case of classical mechanics. If, following the empirical-positivist understanding, a theory must describe observations, then it is easy to show that classical mechanics fails to do so exactly in the same way as QM. Let us explain this. Classical mechanics describes theoretically a `table' in terms of a {\it rigid body}; i.e., a 3-dimensional object in which a number of atoms constitutes the table as a whole unity. The main point we want to stress is that, if we enter a lab in which there is a table, classical mechanics fails to tell us which profile of the table will be observed by us in each instant of time. Classical mechanics describes the table and its evolution ---in case there are forces acting on it---, but remains silent regarding the particular empirical observations made by agents. Classical mechanics simply does not talk about observations or measurements. Instead, it provides a (subject-independent) representation in terms of objects in space-time. If Funes would be taught the positivist understanding of theories and then go into the lab in order to observe the `table' in question, he would certainly not relate the table seen from the front at 3.15 with the table seen from above at 3.20. Observing the `table' from different perspectives Funes would be right to conclude that classical mechanics simply fails to describe what he observed, namely, the profiles of the table. He would be also justified in deriving the following measurement problem for classical mechanics. 

\smallskip 
\smallskip 

\noindent {\it {\bf Classical Measurement Problem (CMP):} Given a specific system of reference, classical mechanics describes mathematically a rigid body in terms of a system constituted in ${\mathcal R}^3$ by a set of different properties. Since classical mechanics predicts the existence of all properties and profiles simultaneously, the question is why do we observe a single profile instead of all the profiles at the same time?}

\smallskip 
\smallskip 

\noindent Classical mechanics simply doesn't tell us which profile we will observe, it only describes the object as a whole. So what is the path from the theoretical description provided by classical mechanics to the profile we observe? From an empirical-positivist there must be also something wrong with Newtonian mechanics since it is not able to account for what we actually observe, namely, the profiles of the table.

\end{document}